\documentclass{easychair}

\usepackage[latin9]{inputenc}
\usepackage[T1]{fontenc}
\usepackage[english]{babel}
\usepackage{amsmath,amsfonts,amssymb}
\usepackage{listings}
\usepackage{url}

\hypersetup{%
pdftitle={Stochastic Formal Methods for Hybrid Systems},%
pdfauthor={Daumas, Lester, Martin-Dorel, Truffert},%
pdfstartview=FitH,%
pdfpagelayout=OneColumn%
}

\newtheorem{theorem}{Theorem}

\newcommand{\ulp}{\text{ulp}}

\newcommand{\R}{\mathbb{R}}

\newcommand{\Typ}{\mathtt{T}} %
\newcommand{\Sig}{\mathcal{S}}
\newcommand{\Prob}{\mathbb{P}}
\newcommand{\Powset}{\mathcal{P}}
\newcommand{\Ex}{\mathbb{E}}
\newcommand{\dP}{~\mathrm{d}\,\Prob}

\newcommand{\To}{\longrightarrow}

\newcommand{\vp}{\vphantom{\displaystyle 2^{2^{2^2}}}}

\begin{document}

\title{Stochastic Formal Methods for Hybrid Systems}

\titlerunning{Stochastic Formal Methods for Hybrid Systems}

\author{%
  Marc Daumas$^1$,  David Lester$^2$, Erik Martin-Dorel$^{1,3}$,  and Annick Truffert$^3$ \\[12pt]
  $^1${\sc eliaus} ({\sc ea 3679 upvd}) and $^3${\sc lamps} ({\sc ea 4217 upvd}) \\
  Perpignan, France 66860,  \href{mailto: marc.daumas@univ-perp.fr; erik.martin-dorel@univ-perp.fr; truffert@univ-perp.fr}{\nolinkurl{{marc.daumas, erik.martin-dorel, truffert}@univ-perp.fr}} \\
  \\
  $^2$School of Computer Science, University of Manchester\\
  Manchester, United Kingdom M13 9PL, \href{mailto:david.r.lester@manchester.ac.uk}{\nolinkurl{david.r.lester@manchester.ac.uk}} \\
}

\authorrunning{Daumas, Lester, Martin-Dorel, Truffert}

\maketitle

\begin{abstract}
  We provide a framework to bound the probability that accumulated
  errors were never above a given threshold on hybrid systems. Such
  systems are used for example to model an aircraft or a nuclear power
  plant on one side and its software on the other side.  This report
  contains simple formulas based on Lévy's and Markov's inequalities
  and it presents a formal theory of random variables with a special
  focus on producing concrete results. We selected four very common
  applications that fit in our framework and cover the common
  practices of hybrid systems that evolve for a long time. We compute
  the number of bits that remain continuously significant in the first
  two applications with a probability of failure around one against a
  billion, where worst case analysis considers that no significant bit
  remains.  We are using PVS as such formal tools force explicit
  statement of all hypotheses and prevent incorrect uses of theorems.
\end{abstract}

\section{Introduction}
Formal proof assistants are used in areas where errors can cause loss
of life or significant financial damage as well as in areas where
common misunderstandings can falsify key assumptions. For this reason,
formal proof assistants have been much used for floating point
arithmetic \cite{Rus98,Har2Ka,BolDau03,DauMel04,DauMel09,DauLesMun08} and
probabilistic or randomized algorithms \cite{Hur02,AudPau06}.
Previous references link to a few projects using proof assistants such
as ACL2 \cite{KauManMoo2K}, HOL \cite{GorMel93}, Coq
\cite{HueKahPau04} and PVS \cite{OwrRusSha92}.

All the above projects that deal with floating point arithmetic aim at
containing worst case behavior. Recent work has shown that worst case
analysis may be meaningless for systems that evolve for a long time as
encountered in the industry.  A good example is a process that adds
numbers in $\pm 2$ with a measure error of $\pm 2^{-24}$.  If this
process adds $2^{25}$ items, then the accumulated error is $\pm 2$,
and note that 10 hours of flight time at operating frequency of 1~kHz
is approximately $2^{25}$ operations.  Yet we easily agree that
provided the individual errors are not correlated, the actual
accumulated errors will continuously be much smaller than $\pm 2$.

We present in Section~\ref{sec/appli} a few examples were this work
can be applied.  We focus on applications for $n$ {\bf counting in
  billions} and a probability of failure about {\bf one against a
  billion}.  Should one of these constraints be removed or lessened,
the problems become much simpler.  The main contribution of this work
is the selection of a few theorems {\bf amenable to formal methods} in
a reasonable time, their application to {\bf software and systems
  reliability}, and our work with PVS.  Section~\ref{sec/stoch}
presents the formal background on probability with Markov's and Lévy's
inequality and how to use this theory to assert software and system
reliability.

Doob-Kolmogorov's inequality was used in previous work
\cite{DauLes07}. It is an application of Doob's inequality that can be
proved with elementary manipulations for second order moment.  It is
better than Lévy's inequality in the sense that it can applied to any
sum of independent and centered variables.  Yet it is limited by
the fact that it bounds only second order moments.

\section{Applications}
\label{sec/appli}

Lévy's inequality works with independent symmetric random variables as
we safely assumed in Sections~\ref{sub/dot} and \ref{sub/iir}.  Doob's
inequality combined with Jensen's one will overcome this restriction
in future formal developments for the applications presented in
Section~\ref{sub/tay} that cannot be treated by Lévy's inequality.
Alas, we foresee that the effort to make Doob's inequality available
in any of the formal tools available today is at least a couple of
years.  Automatic treatment of all the following applications may use
interval arithmetic that has been presented in previous publications
and is now available in formal tools \cite{DauMel04,DauMel09,DauLesMun08}.

\subsection{Long accumulations and dot products}
\label{sub/dot}

A floating point number represents $v = m \times 2^e$ where $e$ is the
exponent, an integer, and $m$ is the mantissa \cite{Gol91}.  IEEE 754
standard \cite{Ste.87} on floating point arithmetic uses
sign-magnitude notation for the mantissa and the first bit $b_0$ of
the mantissa is implicit in most cases ($b_0 = 1$) leading to the
first definition in equation (\ref{eqn/ieee}).  Some circuits such as
the TMS320 \cite{Tex97} use two's complement notation for $m$ leading
to the second definition in equation (\ref{eqn/ieee}).  The sign $s$
and all the $b_i$ are either 0 or 1 (bits).
\begin{equation}
\label{eqn/ieee}
v = (-1)^s \times b_0.b_1 \cdots b_{p-1} \times 2^e
 \qquad \text{or} \qquad
v = (b_0.b_1 \cdots b_{p-1} - 2 \times s) \times 2^e
\end{equation}
In fixed point notation $e$ is a constant provided by the data type
and $b_0$ cannot be forced to 1. We define for any representable
number $v$, the unit in the last place function below, with the
notations of equation~(\ref{eqn/ieee}).
\[
\ulp (v) = 2^{e - p + 1}
\]

The example given in Listing~\ref{lst/int} sums $n$ values. When the
accumulation is performed with floating point arithmetic each
iteration introduces a new round-off error $X_i$.  One might assume
that $X_i$ follows a continuous or discrete uniform distribution on
the range $\pm u$ with $u = \ulp(a_{i})/2$ as trailing digits of
numbers randomly chosen from a logarithmic distribution \cite[pp.
254--264]{Knu97} are approximately uniformly distributed
\cite{FelGoo76}. A significantly different distribution may mean that
the round-off error contains more than trailing digits.

Errors created by operators are discrete and they are not necessarily
distributed uniformly \cite{BusFelGooLin79}. The distribution is very
specific but as soon as we verify that it is symmetric we only have to
bound the moments involved in our main result as in
equation~(\ref{eqn/fund}).
\begin{equation}
\label{eqn/fund}
  \Ex\left(X_i  \right) =   0, ~~~~~
  \Ex\left(X_i^2\right) \le \frac{u^2}{3}, ~~~~~
  \Ex\left(X_i^4\right) \le \frac{u^4}{5}, ~~~~~
  \Ex\left(X_i^6\right) \le \frac{u^6}{7}, ~~~\text{and}~~~
  \Ex\left(X_i^8\right) \le \frac{u^8}{9} %
\end{equation}
If $a_{i}$ uses a directed rounding mode, we introduce $X'_i = X_i -
\Ex(X_i)$ and we use equation (\ref{eqn/fund}) again. We may also
assume that $d_i$ also carries a single error $X_{2, i}$ or a linear
combination of $m-1$ round-off errors $X_{2, i}, \ldots, X_{m, i}$
such that all of them satisfy equation (\ref{eqn/fund}) for a given
$u$.

\begin{lstlisting}[float,caption={Accumulation or dot product},firstnumber=1,label=lst/int]
$a_0 = 0$;
for ($i = 1$; $i <= n$; $i = i + 1$)
  $a_{i} = a_{i-1} + d_i$;
\end{lstlisting}

If $d_i$ is a data obtained by an accurate sensor, we may assume that
the difference between $d_i$ and the actual value $\overline{d_i}$
follows a normal distribution very close to a uniform distribution on
the range $\pm u$ with some new value of $u$. In this cases we model
the error $d_i - \overline{d_i}$ by a symmetric random variable $X_{2,
  i}$ and we use equation (\ref{eqn/fund}).

After $n$ iterations and assuming that all the errors introduced,
$X_i, X_{2, i}, \ldots, X_{m, i}$ are symmetric and independent, we
want the probability that the accumulated errors have exceeded some
user specified bound $\epsilon$:
\begin{equation}
\label{eqn/prob}
  \Prob\left(\max_{1\le i\le n}(|S_i|)\ge\epsilon\right) \le P
  ~~~\text{with}~~~
  S_n = \sum_{i=1}^n \left(X_i + \left\{
    \begin{array}{l l}
      - \Ex(X_i)                & \text{if centering is needed}\\
      \sum_{j = 2}^{m} X_{j, i} & \text{if more variables are needed} \\
      \ldots                    & \ldots
    \end{array}
  \right.\right).
\end{equation}
  
Previous work used Doob-Kolmogorov's inequality.  We will see in
Section~\ref{sec/stoch} that we can exhibit tighter bounds using
Lévy's inequality followed by Markov's one.  Table~\ref{tab/numerical}
present the number of significant bits of the results $\log_2
\epsilon$ for some values of $u$, $n$, $m$, and $P$ in equation
(\ref{eqn/prob}). These values are obtained by using one single value
of $u$, as large as needed. Tighter results can be obtained by using a
specific value of $u$ for each random variable and each iteration.

\begin{table}
  \caption{Number of significant bits with a probability of failure 
  $\displaystyle \Prob\left(\max_{1\le i\le n}(|S_i|)\ge\epsilon\right)$ bounded by $P$} 
  \label{tab/numerical}
  \[
  \begin{array}{| c | c | c | c | c | c | c |} \hline
    u           & n        &  m  & P        & 2k & \epsilon \approx ~~ \text{or} ~~ \epsilon \sim
                                          & \log_2 \epsilon \approx ~~ \text{or} ~~ \log_2 \epsilon \sim \\ \hline
    \vp 2^{-24} & 10^{9}   &  2  & 10^{-9}  & 2  & 68.825        & +6.10     \\ \cline{5-6}
    \vp         &          &     &          & 4  & 0.42832       & -1.22     \\ \cline{5-6}
    \vp         &          &     &          & 6  & 0.085786      & -3.54     \\ \cline{5-6}
    \vp         &          &     &          & 8  & 0.040042      & -4.64     \\ \cline{5-6}
    \vp         &          &     &          & 44 & 0.010153      & -6.62     \\ \hline

    \vp 2^{-24} & 10^{9}   & 10  & 10^{-10} & 2  & 486.66        & +8.92     \\ \cline{5-6}
    \vp         &          &     &          & 4  & 1.7031        & +0.768    \\ \cline{5-6}
    \vp         &          &     &          & 6  & 0.28156       & -1.82     \\ \cline{5-6}
    \vp         &          &     &          & 8  & 0.11939       & -3.06     \\ \cline{5-6}
    \vp         &          &     &          & 48 & 0.023873      & -5.38     \\ \hline

    \vp u       & u^{-3/2} &  1  & u^{3/2}  & 2  & \sqrt[4]{4u^{-2}/9}  & (- \log_2{u} + 1 - \log_2{3}) /2 \\ \cline{5-6}
    \vp         &          &     &          & 4  & \sqrt[8]{4u^{-1}/9}  & (- \log_2{u} + 2 - 2\log_2{3})/8 \\ \cline{5-6}
    \vp         &          &     &          & 6  & \sqrt[12]{100/81}    & (\log_2{10} - 2\log_2{3})     /6 \\ \cline{5-6}
    \vp         &          &     &          & 8  & \sqrt[16]{4900u/729} & (\log_2{u} + 2\log_2{70} - 6\log_2{3}) /16 \\ \hline
  \end{array}
\]
\end{table}

\subsection{Recursive filters operating for a long time}
\label{sub/iir}

Recursive filters are commonly used in digital signal processing and
appear for example in the programs executed by Flight Control Primary
Computers (FCPC) of aircraft. Finite impulse response (FIR) filters
usually involve a few operations and can be treated by worst case
error analysis. However infinite impulse response (IIR) filters may
slowly drift.

Theory of signal processing provides that it is sufficient to study
second order IIR with coefficient $b_1$ and $b_2$ such that polynomial
$X^2 - b_1 X - b_2$ has no zero in $\R$. Listing~\ref{lst/iir}
presents the pseudo-code of one such filter. A real implementation
would involve temporary registers.

\begin{lstlisting}[float=t,caption={Infinite impulse response (IIR) filter},firstnumber=1,label=lst/iir]
$y_{-1} = 0$; $y_0 = d_0$; 
for ($i = 1$; $i <= n$; $i = i + 1$)
  $y_i = d_i - b_1 y_{i-1} - b_2 y_{i-2}$;
\end{lstlisting}

When implemented with fixed or floating point operations, each
iteration introduces a single error $X_i$ or a compound one $X_{1, i}
+ \cdots + X_{m, i}$ in $y_i$.  As these filters are linear, we study
the response to $d_0 = 1$ and $d_i = 0$ otherwise, to deduce the
accumulated effect of all the errors on the output of the filter. It
is defined as the sequence of real numbers such that
\[
y_{-1}  =  0, ~~~~~~~~ 
y_0     =  1, ~~~~~\text{and}~~~~~ 
y_n     =  -b_1 y_{n-1} - b_2 y_{n-2} ~~~~~\text{for all}~~~ n\ge 1.
\]
This sequence can also be defined by the expression
\[
y_n  = b_2^{n/2} ~ \sqrt{b_2 + 2b_1^2} ~ cos \left(\omega_0 + n\omega\right) %
~~~~~\text{with constants}~~ \omega_0 
~~\text{and}~~ \omega.    %
\]
If the filter is bounded-input bounded-output (BIBO) stable, $0 < b2 <
1$ and the accumulated effect of the round-off errors is easily
bounded by $\sqrt{b_2 + 2b_1^2}~/~(1 - \sqrt{b_2})$. Worst case error
analysis is not possible on BIBO unstable systems.  Our work and the
example of Table~\ref{tab/numerical} can be applied to such systems.

\subsection{Long sums of squares and Taylor series expansion of
  programs}
\label{sub/tay}

The previous programs introduce only first order effect of the
round-off errors. We present here systems that involve higher order
errors such as sum of square in Listing~\ref{lst/sos} and power series
of all the random variables as in equation~\ref{eqn/tay}.

Assuming that $d_i$ carries an error $X_i$ in Listing~\ref{lst/sos},
its contribution to the sum of square cannot be assumed to be
symmetric. Lévy's inequality cannot be applied, but Doob's inequality
provides a similar result though it is out of reach with current
formal tools and libraries.

\begin{lstlisting}[float=t,caption={Sum of squares},firstnumber=1,label=lst/sos]
$a_0 = 0$;
for ($i = 1$; $i <= n$; $i = i + 1$)
  $a_{i} = a_{i-1} + d_i*d_i$;
\end{lstlisting}

The output of a system can always be seen as a function $F$ of its
input and its state $(d_0, \ldots, d_n)$. This point of view can be
extended by considering that the output of the system is also a
function of the various round-off errors $(X_0, \ldots, X_q)$
introduced at run-time.  Provided this function can be differentiated
sufficiently, Taylor series expansion provides that
\begin{equation}
\label{eqn/tay}
  \begin{array}{r c l}
    F (d_0, \ldots, d_n, X_0, \ldots, X_q)
    & = & \displaystyle F (d_0, \ldots, d_n, 0,   \ldots, 0)  \\[8pt]
    & + & \displaystyle \sum_{m = 1}^r \frac{1}{m!}
                         \left(\sum_{i = 0}^q X_i \frac{\partial F}{\partial X_i}\right)^{ [m] } (d_0, \ldots, d_n, 0,        \ldots, 0) \\[18pt]
    & + & \displaystyle  \left(\sum_{i = 0}^q X_i \frac{\partial F}{\partial X_i}\right)^{[m+1]} (d_0, \ldots, d_n, \theta_0, \ldots, \theta_q),
  \end{array}
\end{equation}
where $\theta_i$ is between $0$ and $D_i$, and $(\cdot)^{[m]}$ is the
symbolic power defined as
\[   \left(\sum_{i = 0}^q X_i \frac{\partial F}{\partial X_i}\right)^{ [m] }
   = \sum_{0 \le i_1, \ldots, i_m \le q} X_{i_1} \cdots X_{i_m} \frac{\partial^m F}{\partial X_{i_1} \cdots \partial X_{i_m}}.
\]

When the Taylor series is stopped after $m = 2$, we can use Doob's
inequality to provide results similar to the ones presented in
Table~\ref{tab/numerical} provided $X_i$ are symmetric and
independent.  Higher order Taylor series don't necessarily create
sub-martingales but weaker results can be obtained by combining
inequalities on sub-martingales. %

\section{Formal background on probability}
\label{sec/stoch}

\subsection{A generic and formal theory of probability}
\label{sub/prob}

\begin{figure*}
  \begin{center}\fbox{\begin{minipage}{0.98\linewidth}
  {\scriptsize \input{new_proba}}
  \end{minipage}}\end{center}
  \caption{Abbreviated probability space file in PVS}
  \label{lst/prob}
\end{figure*}

We rebuilt the previously published theory of probability spaces
\cite{DauLes07} as a theory of Lebesgue's integration recently became
fully available. The new PVS development in Figure~\ref{lst/prob},
still takes three parameters: $\Typ$, the sample space, $\Sig$, a
$\sigma$-algebra of permitted events, and $\Prob$, a probability
measure, which assigns to each permitted event in $\Sig$, a
probability between $0$ and $1$.  Properties of probability that are
independent of the particular details of $\Typ$, $\Sig$, and $\Prob$
are then provided in this file.

A random variable $X$ is a measurable application from $(\Typ,\Sig)$ to any
other measurable space $(\Typ',\Sig')$.  In most theoretical developments of
probability $\Typ$, $\Sig$, and $\Prob$ remain generic as computations
are carried on $\Typ'$. Results on real random variables use $\Typ' =
\R$ whereas results on random vectors use $\Typ' = \R^n$. Yet both
theories refer explicitly to the Borel sets of $\Typ'$.

As the Borel sets of $\R^n$ are difficult to grasp, most authors
consider finite $\Typ$ and $\Sig=\Powset(\Typ)$ for discrete random
variables in introductory classes.  This simpler analysis is meant
only for educational purposes and most results of probability
considered for formal methods can be implemented with generic $\Typ$,
$\Sig$, and $\Prob$ parameters.

Handling discrete and continuous random variables through different
$\Typ$ and $\Sig$ parameters is not necessary and it is contrary to
most uses of probability spaces in mathematics. Such variables can be
described on the same generic $\Typ$, $\Sig$, and $\Prob$ parameters in
spite of their differences. In practice, we use $\Typ' = \R$ or $\Typ'
= \R^n$, and for discrete variables we can choose countable codomains.

Similarly, many authors work on sections $\{X \le x\}$ rather than
using \emph{the inverse images of Borel sets} of $\Typ'$ because the
latter are difficult to visualize.  Such a simplification is valid
thanks to Dynkin's systems. But using abstract Borel sets rather than
sections in formal methods often leads to easier proofs.

\subsection{A concrete theory of expectation}

The previous theory of random variables \cite{DauLes07} made it
possible to define them and to use and derive their properties. Very
few results were enabling users to actually compute concrete results
on random variables. Most of such results lie on a solid theory of the
expected value. As most theorems in the later theory are corollaries
of a good theory of Lebesgue's integration, we have developed a formal
measure theory based on Lebesgue's integration and we develop formal
theorems on expected values as needed in our applications.

The expected value is the (unique) linear and monotonous operator
$\Ex$ on the set of $\Prob$-integrable random variables that satisfies
Beppo-Lévy's property and such that $\Ex(\chi_A) = \Prob(A)$ for all
$A\in \Sig$. We can also use the following definition when Lebesgue's
integral exists:
\[ \Ex(X) = \int_{\Typ} X \dP. \]

Markov's inequality below is heavily used to obtain concrete
properties on random variables.

\begin{theorem}[Markov's inequality]
  For any random variable $X$ and any constant $\epsilon$,
  \[\Prob\left(|X|\ge\epsilon\right) \le \frac{\Ex\left(|X|\right)}{\epsilon}.\]
\end{theorem}

\bigskip

Many theorems relate to independent random variables and their proof
are much easier once independence is well defined. The family
$\left(X_1, \ldots, X_n\right)$ is independent if and only if, for any
family of Borel sets $(B_1, \ldots, B_n)$,
\[              \Prob\left(\bigcap_{i=1}^n ~ (X_i \in B_i) \right) =
  \prod_{i=1}^n \Prob\left(X_i \in B_i\right).
\]
The following characteristic property is used a lot on families of
independent variables:\\ For any family of Borelean functions 
$(h_1, \ldots, h_n)$ %
such that the $h_i(X_i)$ are $\Prob$-integrable,
\[               \Ex\left(\prod_{i=1}^n h_i(X_i) \right) =
   \prod_{i=1}^n \Ex\left(              h_i(X_i) \right).
\]

It is worth noting that the fact that $n$ random variables are
independent is not equivalent to the fact that any pair of variables
is independent and cannot be built recursively from $n-1$ independent
random variables.

Future work may lead us to implement a theory of the law $\Prob_X$
associated to each random vector $X:\Typ\To\R^n$, with a ``transfer''
theorem for any Borelean function $h:\R^n\To\R$ below and most
properties of Lebesgue's integral including Fubini's theorem.
\[\Ex(h(X)) = \int_{\Typ} h(X) \dP = \int_{\R^n} h {\dP_X}\]

\subsection{Almost certain a priori error bound}
\label{sec/prob}

What we are actually interested in is whether a series of calculations
might accumulate a sufficiently large error to become meaningless. In
the language we have developed, we are computing the probability that
a sequence of $n$ calculations has failed because it has exceeded the
$\epsilon$ error-bound somewhere.

\begin{theorem}[Corollary of Lévy's inequality]
  Provided the $(X_n)$ are independent and symmetric the following
  property holds for any constant $\epsilon$.
  \[\Prob\left(\max_{1\le i\le n}(|S_i|)\ge\epsilon\right) \le
    2\,\Prob\left(|S_n|\ge\epsilon\right)\]
\end{theorem}

\medskip

\begin{proof}
  We use a proof path similar to the one published in \cite{Ber01}. We
  define $S_n^{(j)}$ below with Dirichlet's operator $\delta_P$ that
  is equal to 1 if the predicate holds and 0 otherwise.  As the $X_n$
  are symmetric, the random variables $S_n$ and $S_n^{(j)}$ share the
  same probability density function.
  \[S_n^{(j)} = \sum_{i=1}^n (-1)^{\delta_{i > j}} X_i\]

  We now define $N = \inf \{k ~ \text{such that} ~ |S_k| \ge         %
  \epsilon\}$ with the addition that $\inf \varnothing = +\infty$ and
  similarly $N^{(j)} = \inf \{k ~ \text{such that} ~ |S_k^{(j)}| \ge
  \epsilon\}$. Events $\max_{1\le i\le n}(|S_i|)\ge\epsilon$ and $N
  \le n$ are identical.  Furthermore
  \[              \Prob\left(|S_n      |\ge\epsilon\right)                 = 
    \sum_{j=1}^n \Prob\left(|S_n      |\ge\epsilon ~ \cap ~ N = j\right) = 
    \sum_{j=1}^n \Prob\left(|S_n^{(j)}|\ge\epsilon ~ \cap ~ N = j\right).
  \]

  As soon as $j \le n$, $2S_j = S_n + S_n^{(j)}$ and $2|S_j| = |S_n| +
  |S_n^{(j)}|$. Therefore, the event \(\{|S_j|\ge\epsilon\}\) is included in %
  \(\{|S_n      |\ge\epsilon\} \cup
    \{|S_n^{(j)}|\ge\epsilon\}\)
    and
  \[             \Prob\left(N \le n                              \right) =
    \sum_{j=1}^n \Prob\left(|S_j      |\ge\epsilon ~ \cap ~ N = j\right) \le 
    \sum_{j=1}^n \Prob\left(|S_n      |\ge\epsilon ~ \cap ~ N = j\right)
  + \sum_{j=1}^n \Prob\left(|S_n^{(j)}|\ge\epsilon ~ \cap ~ N = j\right).
  \]

  This ends the proof of Lévy's inequality.
\end{proof}

Should we need to provide some formula beyond the hypotheses of Lévy's
inequality, we may have to prove Doob's original inequality for
martingales and sub-martingales \cite{Nev72} in PVS. It follows a
proof path very different from Doob-Kolmogorov's inequality but it is
not limited to second order moment and it can be applied to any
sub-martingale $S_i^{2k}$ with $k \ge 1$ to lead to
\[\Prob\left(\max_{1\le i\le n}\left(|S_i|\right)\ge\epsilon\right)
   \le \frac{\Ex\left(S_n^{2k}\right)}{\epsilon^{2k}}.
\]

Shall we need to create a sub-martingale different from $S_i^{2k}$,
we may have to prove Jensen's conditional inequality that let us
introduce $h(|S_i|)$ where $h: \R_+ \rightarrow \R_+$ is convex. The
bound becomes $\Ex\left(h\left(|S_n|\right)\right)/h(\epsilon)$.

We use Markov's inequality applied to $S_n^k$ in order to obtain the results of
Table~\ref{tab/numerical}:
\[\Prob\left(|S_n|\ge\epsilon\right) = \Prob\left(|S_n^k|\ge\epsilon^k\right)
  \le \Ex\left(|S_n^k|\right)/\epsilon^k.
\]
Formulas 
\[\begin{array}{r c l}
  \Ex(S_n^2) & = &u^2 \left(
        \tfrac{1}{3} n      \right) \\[8pt]
  \Ex(S_n^4) & = &u^4 \left(
        \tfrac{1}{5} n
      + \tfrac{1}{3} n(n-1) \right) \\[8pt]
  \Ex(S_n^6) & = & u^6 \left(
        \tfrac{1}{7} n
      +              n(n-1)
      + \tfrac{5}{9} n(n-1)(n-2) \right) \\[8pt]
  \Ex(S_n^8) & = & u^8\left(
      \tfrac{1}{9}   n
    + \tfrac{41}{15} n(n-1)
    + \tfrac{14}{3}  n(n-1)(n-2)
    + \tfrac{35}{27} n(n-1)(n-2)(n-3) \right)
\end{array}\]
are based on the binomial formula for independent symmetric random
variables
$$\Ex\left(S_n^{2k}\right) =
  \sum_{k_1 + k_2 + \cdots + k_n = k} (2k)!
  \frac{\Ex\left(X_1^{2k_1}\right)}{(2k_1)!}
  \frac{\Ex\left(X_2^{2k_2}\right)}{(2k_2)!}
  \cdots
  \frac{\Ex\left(X_n^{2k_n}\right)}{(2k_n)!}.
$$

\begin{proof}
  We first prove the formula below by induction on $n$ for any
  exponent $m$.
  \[\Ex\left(S_n^{m}\right) = 
    \sum_{m_1 + m_2 + \cdots + m_n = m} m!
    \frac{\Ex\left(X_1^{m_1}\right)}{m_1!}
    \frac{\Ex\left(X_2^{m_2}\right)}{m_2!}
    \cdots
    \frac{\Ex\left(X_n^{m_n}\right)}{m_n!}
  \]

  It holds for $n = 1$. We now write the following identity based on
  the facts that $X_n$ are independent and symmetric.
  $\Ex\left(S_{n+1}^{m}\right) = \Ex\left(\left(S_n +
      X_{n+1}\right)^{m}\right)$ is also equal to
  \[\Ex\left(\sum_{m_{n+1} = 0}^p \frac{m!}{(m-m_{n+1})!m_{n+1}!} X_{n+1}^{m_{n+1}} S_n^{m-m_{n+1}} \right)
     =
  \sum_{m_{n+1} = 0}^p \frac{m!}{(m-m_{n+1})!m_{n+1}!} \Ex\left(X_{n+1}^{m_{n+1}}\right) \Ex\left(S_n^{m-m_{n+1}}\right)\]

  We expand the terms of the sum for $m_{n+1} = 0, \ldots, p$ and $\sum_{i=1}^n m_i = m-m_{n+1}$
  \[\frac{m!}{(m-m_{n+1})!m_{n+1}!}  \frac{(m-m_{n+1})!}{\prod_{i=1}^{n}m_i!}
    \prod_{i=1}^{n+1} \Ex\left(X_i^{m_i}\right)
  \]

  We end the proof for the even values of $m$ after noticing that
  $\Ex\left(X_i^{2k+1}\right) = 0$ for any $i$ and any $k$ since $X_n$
  are symmetric.
\end{proof}

\section{Perspectives and concluding remarks}

To the best of our knowledge this paper presents the first application
of Lévy's inequality to software and system reliability of very long
processes with an extremely low rate of failure.  Our results allow
any one to develop safe upper limits on the number of operations that
a piece of numeric software should be permitted to undertake. In
addition, we are finishing certification of our results with PVS. The
major restriction lies in the fact that the slow process of proof
checking has forced us to insist that individual errors are symmetric.

At the time we are submitting this work, the bottleneck is the full
certification of more results using PVS proof assistant. Yet this step
is compulsory to provide full certification to future industrial uses.
We anticipate no problem as these results are gathered in textbooks in
computer science and mathematics.  This library and future work will
be included into NASA Langley PVS
library\footnote{\url{http://shemesh.larc.nasa.gov/fm/ftp/larc/PVS-library/pvslib.html}.}
as soon as it becomes stable.

The main contribution of this work is that we selected theorems that
produce significant results for extremely low probabilities of failure
of systems that run for a long time and that are amenable to formal
methods. During our work, we discarded many mathematical methods that
would need too many operations or that would be too technical to be
implemented with existing formal tools.

Notice that this work can be applied to any sequence of independent
and symmetric random variables that satisfy equation (\ref{eqn/fund}).
It is worth pointing out one more time that violating our assumption
(independence of errors) would lead to worse results, so one should
treat the limit we have deduced with caution, should this assumption
not be met.

\section*{Acknowledgment}

This work has been partially funded by CNRS PICS 2533 and by the
EVA-Flo project of the ANR. It was initiated while one of the authors
was an invited professor at the University of Perpignan Via Domitia.

\bibliographystyle{IEEEtran.bst}
\bibliography{alias,perso,groupe,saao,these,livre,arith}

\begin{thebibliography}{10}
\providecommand{\url}[1]{#1}
\csname url@rmstyle\endcsname
\providecommand{\newblock}{\relax}
\providecommand{\bibinfo}[2]{#2}
\providecommand\BIBentrySTDinterwordspacing{\spaceskip=0pt\relax}
\providecommand\BIBentryALTinterwordstretchfactor{4}
\providecommand\BIBentryALTinterwordspacing{\spaceskip=\fontdimen2\font plus
\BIBentryALTinterwordstretchfactor\fontdimen3\font minus
  \fontdimen4\font\relax}
\providecommand\BIBforeignlanguage[2]{{%
\expandafter\ifx\csname l@#1\endcsname\relax
\typeout{** WARNING: IEEEtran.bst: No hyphenation pattern has been}%
\typeout{** loaded for the language `#1'. Using the pattern for}%
\typeout{** the default language instead.}%
\else
\language=\csname l@#1\endcsname
\fi
#2}}

\bibitem{Rus98}
\BIBentryALTinterwordspacing
D.~M. Russinoff, ``A mechanically checked proof of {IEEE} compliance of the
  floating point multiplication, division and square root algorithms of the
  {AMD-K7} processor,'' \emph{LMS Journal of Computation and Mathematics},
  vol.~1, pp. 148--200, 1998. [Online]. Available:
  \url{http://www.onr.com/user/russ/david/k7-div-sqrt.ps}
\BIBentrySTDinterwordspacing

\bibitem{Har2Ka}
\BIBentryALTinterwordspacing
J.~Harrison, ``Formal verification of floating point trigonometric functions,''
  in \emph{Proceedings of the Third International Conference on Formal Methods
  in Computer-Aided Design}, W.~A. Hunt and S.~D. Johnson, Eds., Austin, Texas,
  2000, pp. 217--233. [Online]. Available:
  \url{http://www.springerlink.com/link.asp?id=wxvaqu9wjrgc8l99}
\BIBentrySTDinterwordspacing

\bibitem{BolDau03}
\BIBentryALTinterwordspacing
S.~Boldo and M.~Daumas, ``Representable correcting terms for possibly
  underflowing floating point operations,'' in \emph{Proceedings of the 16th
  Symposium on Computer Arithmetic}, J.-C. Bajard and M.~Schulte, Eds.,
  Santiago de Compostela, Spain, 2003, pp. 79--86. [Online]. Available:
  \url{http://perso.ens-lyon.fr/marc.daumas/SoftArith/BolDau03.pdf}
\BIBentrySTDinterwordspacing

\bibitem{DauMel04}
\BIBentryALTinterwordspacing
M.~Daumas and G.~Melquiond, ``Generating formally certified bounds on values
  and round-off errors,'' in \emph{Real Numbers and Computers}, Dagstuhl,
  Germany, 2004, pp. 55--70. [Online]. Available:
  \url{http://hal.inria.fr/inria-00070739}
\BIBentrySTDinterwordspacing

\bibitem{DauMel09}
\BIBentryALTinterwordspacing
------, ``Certification of bounds on expressions involving rounded operators,''
  \emph{ACM Transactions on Mathematical Software}, 2009, to appear. [Online].
  Available: \url{http://hal.archives-ouvertes.fr/hal-00127769}
\BIBentrySTDinterwordspacing

\bibitem{DauLesMun08}
M.~Daumas, D.~Lester, and C.~Muñoz, ``Verified real number calculations: A
  library for interval arithmetic,'' \emph{IEEE Transactions on Computers},
  2008, to appear.

\bibitem{Hur02}
\BIBentryALTinterwordspacing
J.~Hurd, ``Formal verification of probabilistic algorithms,'' Ph.D.
  dissertation, University of Cambridge, 2002. [Online]. Available:
  \url{http://www.cl.cam.ac.uk/~jeh1004/research/papers/thesis.pdf}
\BIBentrySTDinterwordspacing

\bibitem{AudPau06}
\BIBentryALTinterwordspacing
P.~Audebaud and C.~Paulin-Mohring, ``Proofs of randomized algorithms in
  {C}oq,'' in \emph{Proceedings of the 8th International Conference on
  Mathematics of Program Construction}, T.~Uustalu, Ed., Kuressaare, Estonia,
  2006, pp. 49--68. [Online]. Available:
  \url{http://dx.doi.org/10.1007/11783596_6}
\BIBentrySTDinterwordspacing

\bibitem{KauManMoo2K}
M.~Kaufmann, P.~Manolios, and J.~S. Moore, \emph{Computer-Aided Reasoning: An
  Approach}.\hskip 1em plus 0.5em minus 0.4em\relax Kluwer Academic Publishers,
  2000.

\bibitem{GorMel93}
M.~J.~C. Gordon and T.~F. Melham, Eds., \emph{Introduction to {HOL}: {A}
  theorem proving environment for higher order logic}.\hskip 1em plus 0.5em
  minus 0.4em\relax Cambridge University Press, 1993.

\bibitem{HueKahPau04}
\BIBentryALTinterwordspacing
G.~Huet, G.~Kahn, and C.~Paulin-Mohring, \emph{The {Coq} proof assistant: a
  tutorial: version 8.0}, 2004. [Online]. Available:
  \url{ftp://ftp.inria.fr/INRIA/coq/current/doc/Tutorial.pdf.gz}
\BIBentrySTDinterwordspacing

\bibitem{OwrRusSha92}
\BIBentryALTinterwordspacing
S.~Owre, J.~M. Rushby, and N.~Shankar, ``{PVS}: a prototype verification
  system,'' in \emph{11th International Conference on Automated Deduction},
  D.~Kapur, Ed.\hskip 1em plus 0.5em minus 0.4em\relax Saratoga, New-York:
  Springer-Verlag, 1992, pp. 748--752. [Online]. Available:
  \url{http://pvs.csl.sri.com/papers/cade92-pvs/cade92-pvs.ps}
\BIBentrySTDinterwordspacing

\bibitem{DauLes07}
\BIBentryALTinterwordspacing
M.~Daumas and D.~Lester, ``Stochastic formal methods: an application to
  accuracy of numeric software,'' in \emph{Proceedings of the 40th IEEE Annual
  Hawaii International Conference on System Sciences}, Waikoloa, Hawaii, 2007,
  p. 7~p. [Online]. Available: \url{http://hal.ccsd.cnrs.fr/ccsd-00081413}
\BIBentrySTDinterwordspacing

\bibitem{Gol91}
\BIBentryALTinterwordspacing
D.~Goldberg, ``What every computer scientist should know about floating point
  arithmetic,'' \emph{ACM Computing Surveys}, vol.~23, no.~1, pp. 5--47, 1991.
  [Online]. Available: \url{http://doi.acm.org/10.1145/103162.103163}
\BIBentrySTDinterwordspacing

\bibitem{Ste.87}
D.~Stevenson \emph{et~al.}, ``An {A}merican national standard: {IEEE} standard
  for binary floating point arithmetic,'' \emph{ACM SIGPLAN Notices}, vol.~22,
  no.~2, pp. 9--25, 1987.

\bibitem{Tex97}
\BIBentryALTinterwordspacing
\emph{{TMS320C3x} --- User's guide}, Texas Instruments, 1997. [Online].
  Available: \url{http://www-s.ti.com/sc/psheets/spru031e/spru031e.pdf}
\BIBentrySTDinterwordspacing

\bibitem{Knu97}
D.~E. Knuth, \emph{The Art of Computer Programming: Seminumerical
  Algorithms}.\hskip 1em plus 0.5em minus 0.4em\relax Addison-Wesley, 1997,
  third edition.

\bibitem{FelGoo76}
\BIBentryALTinterwordspacing
A.~Feldstein and R.~Goodman, ``Convergence estimates for the distribution of
  trailing digits,'' \emph{Journal of the ACM}, vol.~23, no.~2, pp. 287--297,
  1976. [Online]. Available: \url{http://doi.acm.org/10.1145/321941.321948}
\BIBentrySTDinterwordspacing

\bibitem{BusFelGooLin79}
\BIBentryALTinterwordspacing
J.~Bustoz, A.~Feldstein, R.~Goodman, and S.~Linnainmaa, ``Improved trailing
  digits estimates applied to optimal computer arithmetic,'' \emph{Journal of
  the ACM}, vol.~26, no.~4, pp. 716 -- 730, 1979. [Online]. Available:
  \url{http://doi.acm.org/10.1145/322154.322162}
\BIBentrySTDinterwordspacing

\bibitem{Ber01}
\BIBentryALTinterwordspacing
J.~Bertoin, ``Probabilités,'' 2001, cours de licence de mathématiques
  appliquées. [Online]. Available:
  \url{http://www.proba.jussieu.fr/cours/bertoin.pdf}
\BIBentrySTDinterwordspacing

\bibitem{Nev72}
J.~Neveu, Ed., \emph{Martingales à temps discret}.\hskip 1em plus 0.5em minus
  0.4em\relax Masson, 1972.

\end{thebibliography}

\end{document}